\documentclass[aps,twocolumn,showpacs,preprintnumbers,amsmath,superscriptaddress]{revtex4}
\usepackage{graphicx} \usepackage{epsf}

\begin{document} 
\def\h{\langle h \rangle}

\title{Simplified Langevin approach to the Peyrard-Bishop-Dauxois model of DNA}

\author{F. de los Santos}
\affiliation{Instituto de~F\'\i sica Te\'orica y Computacional Carlos I  and \\
Departamento de Electromagnetismo y F\'\i sica de la Materia, \\
Facultad de Ciencias, Universidad de Granada, 18071 Granada, Spain}

\author{Omar Al Hammal}
\affiliation{Instituto de~F\'\i sica Te\'orica y Computacional Carlos I  and \\
Departamento de Electromagnetismo y F\'\i sica de la Materia, \\
Facultad de Ciencias, Universidad de Granada, 18071 Granada, Spain}

\author{Miguel A. Mu\~noz}
\affiliation{Instituto de~F\'\i sica Te\'orica y Computacional Carlos I  and \\
Departamento de Electromagnetismo y F\'\i sica de la Materia, \\
Facultad de Ciencias, Universidad de Granada, 18071 Granada, Spain}
\date{\today}

\begin{abstract}
  A simple Langevin approach is used to study stationary properties of
  the Peyrard-Bishop-Dauxois model for DNA, allowing known 
  properties to be recovered in an easy way. Results are shown for the denaturation
  transition in homogeneous samples, for which some implications, so far overlooked,
  of an analogy with equilibrium wetting transitions are
  highlighted. This analogy implies that the order-parameter,
  asymptotically, exhibits a second order transition even if it may be
  very abrupt for non-zero values of the stiffness parameter.  Not
  surprisingly, we also find that for heterogeneous DNA, within this
  model the largest bubbles in the pre-melting stage appear in
  adenine-thymine rich regions, while we suggest the possibility of
  some sort of not strictly local effects owing to the merging of
  bubbles.
\end{abstract}

\maketitle

The DNA thermal melting transition (also called denaturation, coiling,
or un-zipping) occurs when, above a certain critical temperature, the
double-stranded DNA molecule unravels into two separate coils, while
for smaller temperatures (pre-melting stage) only localized openings or
bubbles exist \cite{Wartell}. This phase transition is of importance
for DNA duplication and transcription, and many studies have
scrutinized its nature (whether first or second order), trying to pin
down the relevant traits of the rich phenomenology experimentally observed 
(a nonexhaustive list of references is \cite{PS,Phy,PBD,Varios,Terry,Benham}). Moreover, it has been
suggested that the dynamics of a DNA molecule in its pre-melting stage
may play a role in its own transcription initiation. Indeed, bubbles
are determined by sequence specificity and they have been reported to
occur with high probability in the neighborhood of the, functionally
relevant, {\it transcription start site} (TSS) and near other
regulatory sites, facilitating further microbiological activity
\cite{Benham,Choi,Kalosakas}.

This relation between thermal dynamics and biological functionality
has been claimed to be borne out by experimental data from real
promoter DNA sequences and is supported by results from a theoretical
model (see below) \cite{Choi,Kalosakas}.  Even if this might differ
from biological, protein mediated processes, studies of thermal
properties of the DNA by itself are a first step forward in
understanding more complex situations \cite{Wartell} (see
\cite{Benham2} for a different view).

Let us mention some observations in this context, which have been the
object of recent analyses. Even though one would expect that
adenine-thymine (AT-)rich regions should be more prone to sustain
bubbles than guanine-cytosine-(GC-)rich ones (as AT pairs bind 
the two strands more weakly than GC ones \cite{Wartell}), counterintuitive
situations in which this is not the case have been reported
\cite{GCrich,Benham}.  In the same vein, the dependence of bubble
formation on the specific base-pair sequence was reported to be highly
nonlocal: Upon mutation of two AT base-pairs into two (stronger) GC
base-pairs near the TSS, rendering a specific promoter sequence
completely inactive for transcription, the opening profiles of the
original sequence and its mutant variant differed not only in the
expected suppression of the large thermal opening near the TSS, but
also in a sizable increase in the probability of formation of a bubble at
a distant base pair \cite{Kalosakas}. However, subsequent studies
using more efficient methods for the calculation of bubble statistics
in the  Peyrard-Bishop-Dauxois (PBD) model \cite{Erp,rapti} did not confirm the above non-local
scenario, and pointed to more localized effects. See
\cite{Bishop,Erp2} for recent developments on this interesting
problem.

Many of these and other relevant issues have been investigated by
employing the PBD model \cite{PBD} (see
below). The model phenomenology has been profusely analyzed by means
of various analytical and numerical techniques: transfer integral
calculations, Monte Carlo simulations, molecular dynamics, and
Langevin dynamics, and the results have been found to properly describe 
experiments on the melting transition \cite{Campa}, pre-melting
bubbles \cite{Bishop}, etc. Let us caution that under certain
circumstances, torsional effects (absent in the PBD model) should be
included to properly account for some of the described phenomenology
\cite{Benham,Benham2,Barbi}.

In this Brief Report we reconsider the DNA thermal denaturation problem
analyzing the PBD model \cite{PBD} by means of a different, simplified
Langevin approach. This strategy allows us to: (i) reproduce
numerically in a relatively easy way the stationary bubble probability
distribution and other statistical properties for both homogeneous and
heterogeneous sequences;
(ii) establish an analogy with well-known {\it equilibrium wetting}
problems, deeper than previously thought, permitting us to infer results
about the order of the denaturation transition.

In the PBD model the stretching of hydrogen-bonds between
corresponding base-pairs is represented by a set of continuous
variables $\{h_n\}$ (at positions $n=1, ...,N$ where $N$ is the chain
length). The model is defined by the following Hamiltonian
\cite{PBD}
\begin{equation}
H=\sum_{n=1}^N \left(
\frac{1}{2} m{\dot h}_n^2+V(h_n)+W(h_n,h_{n-1})
\right).
\label{hamiltonian}
\end{equation}
The first term is the kinetic energy for bases of mass $m$.  The
second one stands for the interaction between opposite bases as
described by the Morse potential
\begin{equation}
V(h_n)= D_n (e^{- a_n h_{n}}-1)^2,
\label{morse}
\end{equation}
where $D_n$ is the dissociation energy of the $n$th base pair and
$a_n$ denotes the spatial range of the potential. Standard,
empirically found pair-base-dependent parameter values are customarily
employed: $D_n({\rm AT})=0.05$ eV, $D_{n}({\rm GC})=0.075$ eV, $a_n({\rm AT})=4.2$ \AA$^{-1}$, 
and $a_n({\rm GC})=6.9$ \AA$^{-1}$
\cite{Campa}.  Finally, the third {\em stacking} term arises from the
interaction between adjacent bases along the DNA molecule
\cite{PBD}.  It reads
\begin{equation}
 W(h_n,h_{n-1})= \frac{k}{2} \left( 1+ \rho
 e^{-\alpha(h_n+h_{n-1})}\right)(h_n-h_{n-1})^2,
\label{stackPB}
\end{equation}
where the values of $k$, $\rho$, and $\alpha$ are determined from
fittings of experimental DNA denaturation curves \cite{Campa}:
$k=0.025$ eV\AA$^2$, $\rho=2$, $\alpha=0.35$ \AA$^{-1}$. The
nonvanishing {\it stiffness parameter} $\rho$ captures the fact that
the double-stranded backbone is more rigid than the unwound strands
(controlled by a standard elastic interaction). Note that this model
includes only transverse degrees-of-freedom for nucleotides.

The average stretching at each site $\langle h_n \rangle$ and its
space-averaged counterpart $\langle h \rangle$, as well as 
$\langle e^{-h} \rangle$, which can be interpreted as the density of 
closed base-pairs, are the standard order-parameters.

Different scenarios have been reported for the denaturation transition
depending on the stiffness parameter $\rho$ and the randomness of the
DNA sample. In the simplest case $\rho=0$ \cite{PBD}, the stacking term
is harmonic and a smooth (second-order) denaturation transition is
known to occur for both homogeneous and heterogeneous DNA
\cite{Terry,Theo}.  On the contrary, non-vanishing $\rho$ and
heterogeneous sequences lead to very abrupt thermal denaturation
curves that exhibit a multistep behavior in line with experimental
observations \cite{Terry}.

The case of nonzero $\rho$ and homogeneous DNA is still unsettled as
the transition has been reported to be (i) first-order-like yet with
a diverging correlation length in \cite{Theo,Joyeux} and (ii)
second order although very sharp in appearance \cite{Terry}.  We shall
return to this issue below.  Let us also remark that, as pointed out
in \cite{Theo}, a continuous transition for the order parameter
$\langle h \rangle$ with associated critical exponents and a diverging
length scale could be compatible (if $\rho \neq 0$) with the
number of bound pairs $\langle n \rangle$ exhibiting a discontinuity
at the transition.

In evaluating the partition function associated with the Hamiltonian
Eq.(\ref{hamiltonian}), the kinetic terms factorize and, as a result,
can be dropped out if the focus is only on equilibrium configurational
properties. In such a case, the equilibrium state can be recovered
from the configurational part $H'$ of $H$ (including only $V$ and $W$
terms) and, therefore, can be reproduced from the stationary solution
of the associated Langevin equation,
\begin{equation}
\frac{\partial h_n(r,t)} {\partial t}= -\frac{\partial H'(h_n)}{\partial h_{n}} 
+\sigma \eta(r,t),
\label{langevinpbd}
\end{equation}
where $\eta$ is a Gaussian white noise and $\sigma$ its amplitude.  In
the following, Eq.(\ref{langevinpbd}) is taken as the starting point for
study, and an Euler algorithm is used to solve it. This differs from
previous Langevin studies in that inertial terms do not appear,
enabling slightly faster computational studies.  A similar approach
was used in \cite{Anxo}. Let us stress that the dynamics imposed by
Eq.(\ref{langevinpbd}) is a fictitious one, not related to real DNA
dynamics (which is not purely relaxational), but leads to the same
stationary probability distribution as the original one.

{\em Homogeneous DNA}. We begin by studying the case of homogeneous
samples with only GC base pairs. The temperature $T$ is the control
parameter, and the value of $\sigma$ is obtained from the
fluctuation-dissipation relation.  We have run simulations in systems
of size $2^{17}$, initializing all the base-pairs to $h(t=0)=2$ and
letting them evolve until a stationary state is reached. $\h$ was
monitored as a function of time for zero and nonzero values of $\rho$.
At low temperatures $\h$ saturates to a finite value whereas at high
enough temperatures it diverges as $t^{1/4}$ (see below),
signaling a phase transition. While for
$\rho=0$ a smooth (continuous) transition is observed, for $\rho=2$ it
is rather abrupt (results not shown), being apparently first order. The
same picture, in line with previous numerical results \cite{PBD}, can
also be drawn by monitoring $\langle e^{-h} \rangle$, but our results are
not fully conclusive.

As originally argued in \cite{Terry}, as $h_n \approx h_{n-1}$, the
exponential factor in Eq.(\ref{stackPB}) can be approximated by 
$e^{-\alpha h_n}$ without provoking any significant effect. 
If $\rho=0$, $H'$
is readily recognized (apart from constant terms) as a discretized
version of the continuous Hamiltonian
\begin{equation}
H_{ew}=\int dx \left(\frac{k}{2} (\nabla h)^2 + w_1 e^{-ah} +w_2 e^{-2ah} \right),
\label{wetting_hamiltonian}
\end{equation}
where $w_1$, $w_2$, and $k$ are generic parameters. $H_{ew}$ is the
standard interfacial Hamiltonian for equilibrium critical wetting
transitions in the presence of short-ranged forces, i.e. the unbinding
of the interface separating two coexisting phases from a wall, which
occurs upon increasing the temperature \cite{review_wetting}.
At this point, we recall that in wetting phenomena continuum models are valid
approximations to lattice models as long as $T$ is above the roughening
temperature $T_R$, which is $T_R=0$ in $d=1$ ($d=2$ bulk).

Although the connection between wetting and DNA denaturation has
already been recognized (see, for instance, \cite{Terry,Anxo}) some of
its consequences have not been fully appreciated. For instance, the
set of recently reported \cite{Theo} critical exponents characterizing
the DNA denaturation transition in the homogeneous cas, $\h \sim
|\delta|^{-\beta}$ and $\xi \sim |\delta|^{-\nu}$ [where
$\delta=(T-T_c)/T_c$], with $\beta=-1$, $\xi$ the correlation length,
and $\nu=2$, are nothing but the two-dimensional critical wetting
exponents dating back to the early 1980s \cite{review_wetting}.
Furthermore, the density of closed base pairs scales as
$\langle h^{-1} \rangle \sim |\delta|$ (see \cite{Terry}), as
corresponds to the surface order parameter in a wetting context
\cite{review_wetting}. Additionally, since in equilibrium wetting the
dynamic critical exponent $z$, defined by $\xi \sim t^{1/z}$, is $z=2$,
the thickness of the wetting layer grows as $t^{1/4}$ \cite{lipowsky},
in agreement with the value reported above for the PBD model.  To the
best of our knowledge, these correspondences have not been established
before.

More interestingly, the implications of the wetting
analogy can be extended to the nonzero-$\rho$ case. In the wetting context,
a long-standing problem, regarding the order of the transition in
three-dimensional systems, has been recently solved \cite{parry}. The
original renormalization-group calculations led
to the prediction of non-universal results in blatant disagreement
with computational studies \cite{binder} and experiments \cite{bonn},
both of which yield a mean-field-like second-order phase transition.
An early attempt to reconcile theory and experiments questioned the
validity of the effective Hamiltonian Eq.(\ref{wetting_hamiltonian})
to describe equilibrium wetting and concluded that $k$ in
Eq.(\ref{wetting_hamiltonian}) should be replaced by a
position-dependent {\it stiffness coefficient} $k(h)=k+w'_1e^{-\alpha h} +
w'_2ah e^{-2\alpha h}+ \cdots$ \cite{jf}. Curiously enough, with
only the leading correction included in $k(h)$, this Hamiltonian is the
continuous counterpart of the PBD one.

In critical wetting the parameter $w'_1$ vanishes at the transition
point and, according to a linear renormalization-group study, only the
term proportional to $w'_2$ is capable of destabilizing the critical
wetting transition, driving the transition weakly first-order
in $d=3$ \cite{jf}. A subsequent
investigation allowed the analysis to be extended, with the conclusion that a
first-order transition can appear only for dimensions $d\gtrsim 2.41$
\cite{boulter}. Remarkably,  it has been shown \cite{parry}
that by including the whole series expansion the experimental and
computational results can be finally reproduced.

These results can be adapted for homogeneous DNA melting. Indeed,
by switching on a nonvanishing $w'_1$ and truncating the series to first
order, we do not expect the above conclusions to change qualitatively,
since it is naively expected that $w'_1$ plays a similar role to
$w'_2$ (the detailed proof of this is not straightforward and will be
published elsewhere).  Therefore, using the wetting analogy, the
one-dimensional melting transition for homogeneous DNA sequences
should be {\it asymptotically} continuous for $\langle h \rangle$, in
agreement with some previous transfer integral analyses \cite{Terry},
but in partial disagreement with other calculations \cite{Theo,Joyeux}.
Reconciling all these results remains an open challenging task.

Our conclusion about the order of the transition might change if we consider
versions of the PBD model embedded in a three-dimensional
space \cite{Barbi} where bubble entropic effects are expected to play
a crucial role \cite{Phy}. Note also that for such three-dimensional
models the analogy with wetting problems breaks down.

{\em Heterogeneous DNA}.
Following
the recent literature, we have simulated our model for two particular
sequences of $69$ base-pairs: the adeno-associated viral P5
(AAVP5) promoter and a mutation of it inactive for transcription \cite{Choi}. 
In the mutant sequence two AT bases located near the TSS
at positions $48$ and $49$ are replaced by (more tightly bound) GC
base pairs. In our analyses a bubble is defined as a group of adjacent
sites that satisfies the condition $h>1.5$.
To avoid finite-size effects, we use periodic boundary conditions on
lattices of sizes $L=690$ and 6900 consisting of $10$ and $100$
replicas, respectively, of the same AAVP5 sequence, After sufficient
ensemble averaging, indistinguishable long-time results are obtained
for both sizes.
\begin{figure}[ht!]
\includegraphics[width=86mm,clip]{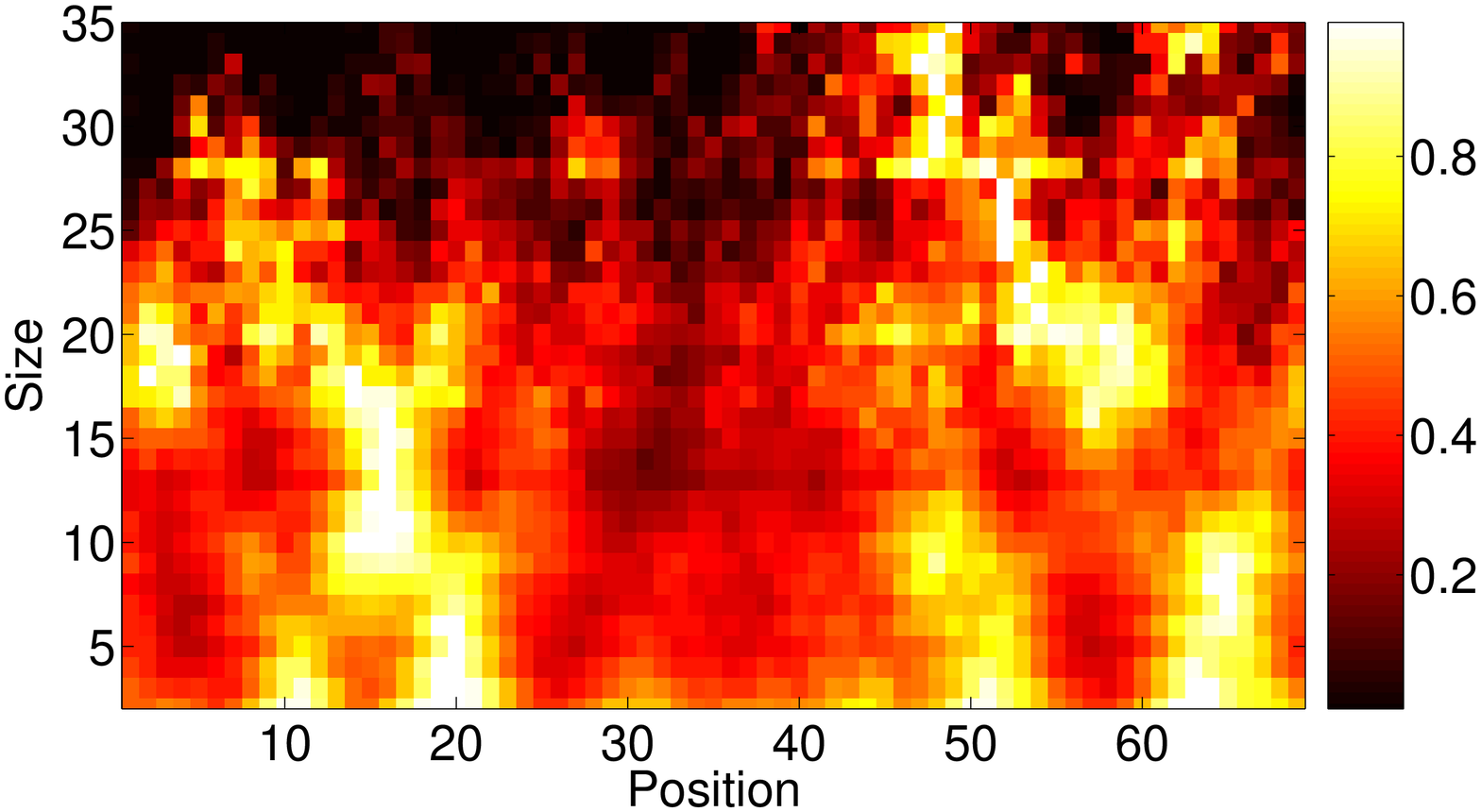}
\includegraphics[width=86mm,clip]{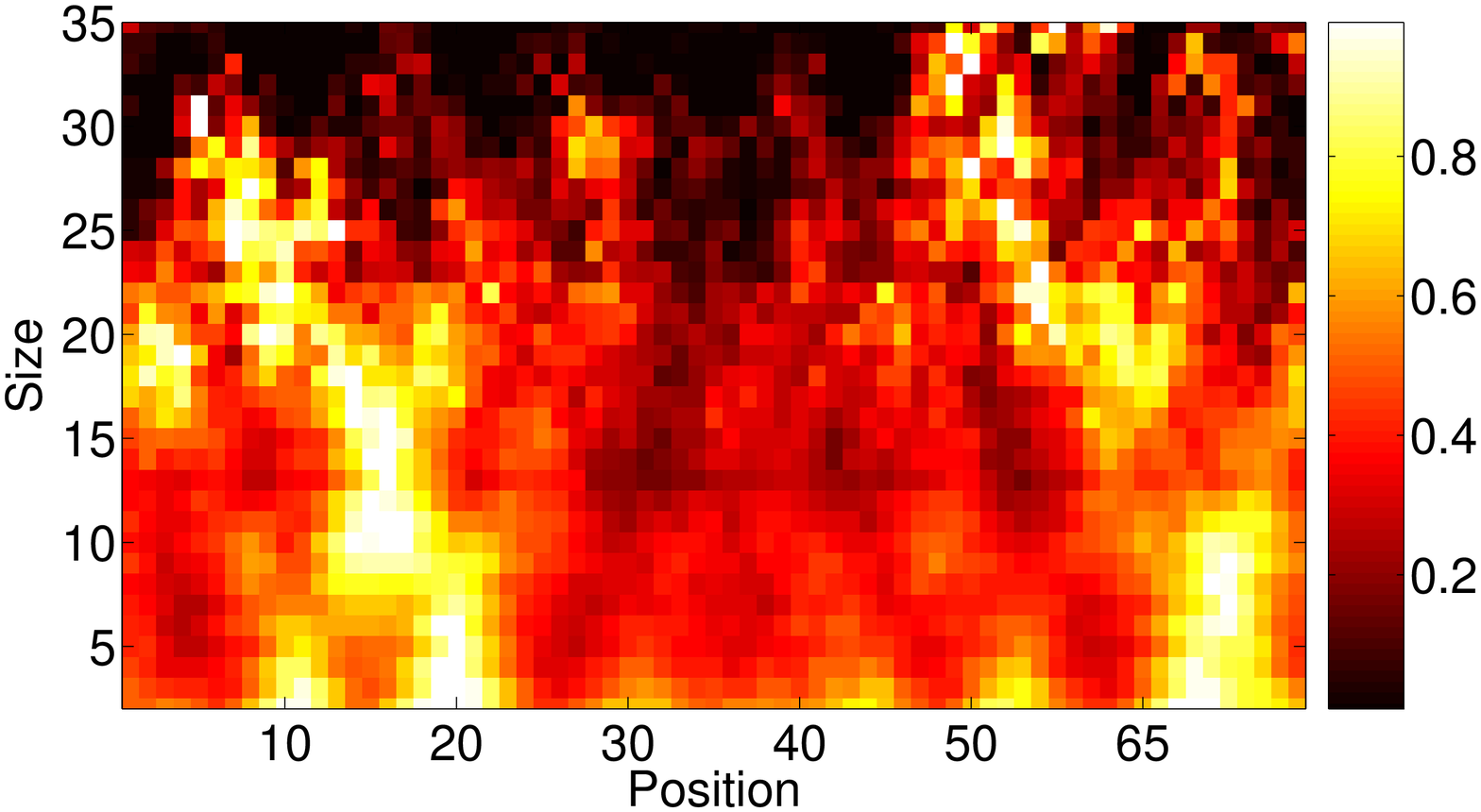}
\caption{(Color online). Probability of bubble opening as a function of position and bubble
size for the AAVP5 promoter (top panel) and the mutant P5 promoter
(bottom panel) at $T=310$K. Probabilities in each row are normalized
to their maximum value as in \cite{Erp}. The results are very similar to
those in \cite{Erp}.}
\label{dna-bubbles1}
\end{figure}
The bubble distributions for the AAVP5 sequence and its mutant are
shown in Fig.\ref{dna-bubbles1}.  It can be seen that the large
bubbles forming around the TSS (top panel) are suppressed in the
mutant sequence (bottom panel) in agreement with experimental
observations \cite{Choi}. The effect of the mutation is quite
local, in line with that obtained in \cite{Erp} and in contrast
to the first claims \cite{Choi}. Observe, also, that bubbles in the
DNA sequence form more frequently where AT bases are more abundant,
as naively expected \cite{Bishop,Erp2}. Situations in which this is not the
case (like those reported in \cite{GCrich}) are likely to be physically
ascribable to torsional effects \cite{Benham,Benham2}. Our conclusion
is that the local bubble-opening probability within the PBD model is
controlled by the relative density of AT base-pairs, in accordance
with \cite{Bishop,Erp2}.

\begin{figure}[ht]
\includegraphics[width=80mm,clip]{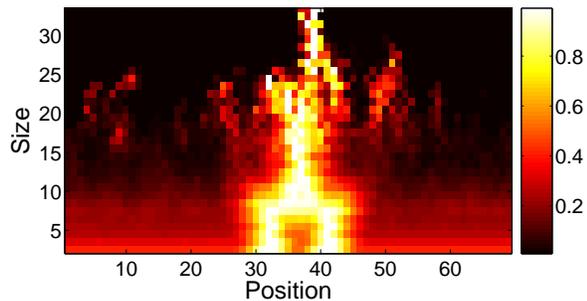}
\caption{(Color online). Bubble merging over a GC region from the openings above two small AT regions.}
\label{dna-bubbles2}
\end{figure}

To explore the possibility of having some sort of nonlocal effect in
bubble formation within the present model, consider an artificial
chain with a GC-rich region separating two AT-rich zones (see
Fig.\ref{dna-bubbles2}).  Small bubbles formed in the two AT-rich
regions might eventually merge together, bridging across the GC region
as illustrated in Fig.\ref{dna-bubbles2}. This can induce the largest
possible bubble to be centered around a GC-rich zone, and nonstrictly
local effects could be generated upon introducing mutations. Further
research is needed to quantify this mechanism and to assess if it is
capable of inducing nonlocal effects by repetition of the above
scenario, which has already been discussed in the literature in
various forms \cite{Novotny}.

In summary, the simple Langevin equation (\ref{langevinpbd}) gives
relatively quick access to the stationary properties of the PBD model
for DNA denaturation. It reproduces many known results for the
homogeneous case, {\em e.g.}, for $\rho=0$ a continuous transition is
obtained. Moreover, we have pointed out that the (recently obtained)
critical exponents are well known for the wetting problem. The analogy
with equilibrium critical wetting can be extended using very recent
developments to the $\rho \neq 0$ case, where also a continuous
transition is predicted (even if it might be a very abrupt one
\cite{Terry,Theo}).  We have also employed the Langevin approach to
study the bubble statistics in heterogeneous real sequences,
confirming the tendency for creation of thermal openings around AT-rich
regions. According to our observations mutations modify the statistics
of bubbles only in a local way. However, nonstrictly-local effects
due to the merging of bubbles could induce large openings in locally
GC-rich regions.

It is our hope that this simple Langevin approach will be useful to
elucidate other aspects of this fascinating field.  
%In particular, the
%approach could be extended to analyze the version of the model with
%torsion effects \cite{Barbi}.

We acknowledge financial support from the Spanish MEyC-FEDER, Project
No. FIS2005-00791 and from Junta de Andaluc{\'\i}a as group FQM-165.


\begin{thebibliography}{99}

\bibitem{Wartell} 
R.M. Wartell and A.S. Benight, 
Phys. Rep. {\bf 126}, 67 (1985), 
and references therein.

\bibitem{PS} 
D. Poland and H.A. Scheraga, 
J. Chem. Phys. \textbf{45}, 1456 (1966).

\bibitem{Phy} 
Y. Kafri, D. Mukamel, and L. Peliti, 
Phys. Rev. Lett. {\bf 85}, 4988 (2000); 
Eur. Phys. J. B. {\bf 27}, 135 (2002);
E. Carlon, E. Orlandini, and A. Stella,
% Roles of stiffness and excluded volume in DNA denaturation
Phys. Rev. Lett. {\bf 88}, 198101 (2002);  
M.S. Causo, B. Coluzzi, and P. Grassberger, 
Phys. Rev. E {\bf 62}, 3958 (2000);  
S. Cocco and R. Monasson, 
Phys. Rev. Lett. {\bf 83}, 5178 (1999).


\bibitem{PBD} T. Dauxois, M. Peyrard, and A.R. Bishop,
Phys. Rev. E, \textbf{47}, R44 (1993). 
See also, M. Peyrard and A.R. Bishop, Phys. Rev. Lett. \textbf{62}, 2755 (1989);  
M. Peyrard, Nonlinearity {\bf17}, R1 (2004).


\bibitem{Varios} 
%\bibitem{Hugues}
L.-H. Tang and H. Chat\'e, Phys. Rev. Lett. {\bf 86}, 830 (2001);
%\bibitem{Hwa2}
T. Hwa {\it et al.},
% E. Marinari, K. Sneppen, and L.-H. Tang,
Proc. Natl. Acad. Sci. USA. {\bf 100}, 4411 (2003);
%\bibitem{Zocchi}
V. Ivanov, Y. Zeng, and G. Zocchi, 
Phys. Rev. E {\bf 70}, 051907 (2004);
%V. Ivanov, D. Piontkovski, and G. Zocchi, Phys. Rev. E {\bf 71}, 041909 (2005).
%\bibitem{Azbel} 
M.Y. Azbel,  Phys. Rev. A {\bf 20}, 1671 (1979); 
Phys. Rev. E {\bf 68}, 050901(R) (2003).

\bibitem{Terry} 
D. Cule, and T. Hwa, 
Phys. Rev. Lett. {\bf 79}, 2375 (1997).

\bibitem{Benham}
C. J. Benham, 
Proc. Natl. Acad. Sci. USA {\bf 90}, 2999 (1993);
%J. Mol. Biol. {\bf 255}, 425 (1996).  
R. M. Fye and C. J. Benham,
%Exact Method for Numerically
%Analyzing a Model of Local Denaturation in Superhelically Stressed DNA, 
Phys. Rev. E {\bf 59}, 3408 (1999).
%Benham, C.J. (2001) Stress-Induced DNA Duplex Destabilization in
%Transcriptional Regulation, Proceedings of the 2001 Pacific Symposium
%on Biocomputing (refereed), World Scientific, 103-114.

\bibitem{Choi} 
C. H. Choi {\it et al.}, 
%G. Kalosakas, K. O. Rasmussen,
%M. Hiromura, A. R. Bishop, and A. Usheva, 
Nucleic Acids Res. {\bf 32}, 1584 (2004);
C. H. Choi {\it et al.},
Phys. Rev. Lett. {\bf 96}, 239801 (2006).
%Structurally specific thermal fluctuations identify functional sites for DNA transcription
%\bibitem{bubble_stat3}
%Comment on "Can One Predict DNA Transcription Start Sites by Studying Bubbles?"

\bibitem{Kalosakas} 
G. Kalosakas {\it et al.},   
%K. \O. Rasmussen, A. R. Bishop, C. H. Choi, and A. Usheva, 
Europhys. Lett. {\bf 68}, 127 (2004).

\bibitem{Benham2} 
C. J. Benham and R.R.P. Singh, 
Phys. Rev. Lett. {\bf 97}, 059801 (2006).

\bibitem{GCrich} 
U. Dornberger, M. Leijon, and H. Fritzsche, 
J. Biol. Chem. {\bf 274}, 6957 (1999).

\bibitem{Erp} 
T.S. van Erp, 
%{\it et al.},
S. Cuesta-L\'opez, J.-G. Hagmann, and M. Peyrard,
Phys. Rev. Lett. {\bf 95}, 218104 (2005); ibid. {\bf 96}, 239802
(2006); ibid. {\bf 97}, 059802 (2006).
%\bibitem{bubble_stat4} T. S. van Erp, S. Cuesta-Lopez,
%J.-G. Hagmann, and M. Peyrard, Phys. Rev. Lett. 
%Reply to the previous comment
%\bibitem{bubble_stat6} T. S. van Erp, S. Cuesta-Lopez,
%J.-G. Hagmann, and M. Peyrard,
% reply to second comment

\bibitem{rapti}
Z. Rapti {\it et al.},
Europhys. Lett. {\bf 74}, 540 (2006).

\bibitem{Erp2}
T.S. van Erp, S. Cuesta-L\'opez, and M. Peyrard, 
Eur. Phys. J. E {\bf 20}, 421 (2006).

\bibitem{Bishop}
S. Ares {\it et al.},
% N. K. Voulgarakis, K. O. Rasmussen and A. R. Bishop:
%Bubble Nucleation and Cooperativity in DNA Melting, 
Phys. Rev. Lett. {\bf 94}, 035504 (2005);
Z. Rapti {\it et al.},
% A. Smerzi, K. \O. Rasmussen, A.R. Bishop, C.H. Choi, and A. Usheva,
Phys. Rev. E {\bf 73}, 051902 (2006);
B.S. Alexandrov {\it et al.}
%L. T. Wille, K. \O. Rasmussen, A. R. Bishop and K. B. Blagoev, 
Phys. Rev. E {\bf 74}, 050901(R) (2006); 
S. Ares and G. Kalosakas, Nano Lett. {\bf 7}, 307 (2007).

\bibitem{Campa} 
A. Campa and A. Giansanti, 
Phys. Rev. E. {\bf 58}, 3585 (1998).
%See also, S. Zdravkovi\'c and M. Satari\'c, Phys. Rev. E {\bf 73}, 021905
%(2006).

\bibitem{Barbi} 
M. Barbi, S. Cocco, and M. Peyrard, 
Phys. Lett. A {\bf 253}, 358 (1999).
%\bibitem{3D} G. F. Calvo and R. F.  Alvarez-Estrada
%Title: Three-dimensional models for homogeneous DNA near denaturation 
%JOURNAL OF PHYSICS-CONDENSED MATTER 17 (50): 7755-7781 DEC 21 2005. 

\bibitem{Theo}
N. Theodorakopoulos, T. Dauxois, and M. Peyrard, 
Phys. Rev. Lett. {\bf 85}, 6 (2000).

\bibitem{Joyeux} S. Buyukdagli and M. Joyeux, Phys. Rev. E {\bf 73},
051910 (2006). M. Joyeux and  S. Buyukdagli,  Phys. Rev. E {\bf 72},
051902 (2005).

\bibitem{Anxo} 
S. Ares and A. S\'anchez,
%Modelling Disorder: the Cases of Wetting and DNA Denaturation, The 
Eur. Phys. J. B {\bf 56}, 253 (2007). 
 
\bibitem{review_wetting}
S. Dietrich, in {\em Phase Transitions and Critical Phenomena},
vol. 12, edited by C. Domb and J. Lebowitz (Academic Press, New York,
1988); M. Schick, in {\em Liquids at Interfaces}, Les
Houches Summer School Proceedings 48, ed. by J. Charvolin,
J.F. Joanny, and J. Zinn-Justin (North Holland, Amsterdam, 1990).

\bibitem{lipowsky}
R. Lipowsky,
J. Phys. A: Math. Gen. {\bf 18}, L585 (1985).

\bibitem{parry}
A.O. Parry, J.M. Romero-Enrique, and A. Lazarides,
Phys. Rev. Lett. {\bf 93}, 086104 (2004).

\bibitem{binder}
K. Binder, D.P. Landau, and D.M. Kroll, 
Phys. Rev. Lett. {\bf 56}, 2272 (1986).

\bibitem{bonn}
D. Ross, D. Bonn, and J. Meunier, 
Nature {\bf 400}, 737 (1999).

\bibitem{jf}
A.J. Jin and M.E. Fisher, 
Phys. Rev. B {\bf 48} 2642 (1993).

\bibitem{boulter}
C.J. Boulter, 
Phys. Rev. Lett. {\bf 79}, 1897 (1997).
%Finally, including all the terms in the expansion Parry et al. \cite{parry}
%recovered the experimental results \cite{binder,bonn}.


%\bibitem{Zeng} 
%Y. Zeng, A. Montrichok, and G. Zocchi,
%Phys. Rev. Lett. {\bf 91}, 148101 (2003); 
%J. Mol. Biol. {\bf 339}, 67 (2004).

%  Bubble dynamics in dsDNA:
% G. Altan-Bonnet, A. Libchaler, and O. Krichevsky,
%Phys. Rev. Lett. 90, 138101 (2003). 

%\bibitem{Klein} A. Santos and W. Klein, arXiv:cond-mat/0610484.

\bibitem{Novotny} 
T. Novotny {\it et al.},
%J. N. Pedersen...
%Bubble merging
Europhys. Lett. {\bf 77}, 48001 (2007).



\end{thebibliography}
\end{document}